\documentclass[aps, preprint, groupedaddress, floatfix,nofootinbib,longbibliography]{revtex4-1}

\usepackage{dutchcal}
\usepackage{epsfig}
\usepackage{graphicx}
\usepackage{amsmath}
\usepackage{mathtools}
\usepackage{bbold}
\usepackage{tabularx, booktabs}
\usepackage[normalem]{ulem}
\usepackage{soul}  %ywfang, strikethrough font
\usepackage{color} % color font by ywfang
\usepackage{multirow} %multirow for table by ywfang
\usepackage{xr}
\begin{document}

\title{A substantial
  hybridization between correlated Ni-$d$ orbital and
  itinerant electrons in infinite-layer nickelates}

\author{Yuhao Gu$^{1,2,4}$} \author{Sichen Zhu$^{2,3}$}
\author{Xiaoxuan Wang$^{2,3}$} \author{Jiangping Hu$^1$}
\author{Hanghui Chen$^{2,3}$} \affiliation{ $^1$Institute of Physics,
  Chinese Academy of Science, Beijing 100190, China\\ $^2$NYU-ECNU Institute
  of Physics, NYU Shanghai, Shanghai 200122, China\\ $^3$Department of
  Physics, New York University, New York, New York 10003,
  USA\\ $^4$Beijing National Laboratory for Molecular Sciences, State
  Key Laboratory of Rare Earth Materials Chemistry and Applications,
  Institute of Theoretical and Computational Chemistry, College of
  Chemistry and Molecular Engineering, Peking University, 
  Beijing 100871, China } \date{\today}

\begin{abstract}

The discovery of unconventional superconductivity in hole doped
NdNiO$_2$, similar to CaCuO$_2$, has received enormous
attention. However, different from CaCuO$_2$, $R$NiO$_2$ ($R$ = Nd,
La) has itinerant electrons in the rare-earth spacer layer. Previous
studies show that the hybridization between Ni-$d_{x^2-y^2}$ and
rare-earth-$d$ orbitals is very weak and thus $R$NiO$_2$ is still a
promising analog of CaCuO$_2$. Here, we perform first-principles
calculations to show that the hybridization between Ni-$d_{x^2-y^2}$
orbital and itinerant electrons in $R$NiO$_2$ is substantially
stronger than previously thought. The dominant hybridization comes
from an interstitial-$s$ orbital rather than rare-earth-$d$ orbitals,
due to a large inter-cell hopping. Because of the hybridization, Ni
local moment is screened by itinerant electrons and the critical
$U_{\textrm{Ni}}$ for long-range magnetic ordering is increased. Our
work shows that the electronic structure of $R$NiO$_2$ is distinct
from CaCuO$_2$, implying that the observed superconductivity in
infinite-layer nickelates does not emerge from a doped Mott insulator.
\end{abstract}

\maketitle

\section*{Introduction}

Since the discovery of high-temperature superconductivity in
cuprates~\cite{bednorz1986possible}, people have been attempting to
search for superconductivity in other materials whose crystal and
electronic structures are similar to those of
cuprates~\cite{anisimov1999electronic,hu2016identifying}. One of the
obvious candidates is La$_2$NiO$_4$ which is iso-structural to
La$_2$CuO$_4$ and Ni is the nearest neighbor of Cu in the periodic
table. However, superconductivity has not been observed in doped
La$_2$NiO$_4$~\cite{rao1989superconductivity}.  This is in part due to
the fact that in La$_2$NiO$_4$, two Ni-$e_g$ orbitals are active at
the Fermi level, while in La$_2$CuO$_4$ only Cu-$d_{x^2-y^2}$ appears
at the Fermi level. Based on this argument, a series of nickelates and
nickelate heterostructures have been proposed with the aim of realizing a
single orbital Fermi surface in nickelates. Those attempts started
from infinite-layer
nickelates~\cite{hayward1999sodium,anisimov1999electronic,lee2004infinite},
to LaNiO$_3$/LaAlO$_3$
superlattices~\cite{freeland2011orbital,LNOLMO_superlattice,Hansmann_heterostructure,Han_DMFT_superlattice},
to tri-component nickelate
heterostructures~\cite{chen2013modifying,Disa_heterostructure} and to
reduced Ruddlesden-Popper
series~\cite{botana2017electron,zhang2017large}. Eventually,
superconductivity with a transition temperature of about 15 K has
recently been discovered in hole doped infinite-layer nickelate
NdNiO$_2$~\cite{li2019superconductivity}, injecting new vitality into
the field of high-$T_c$
superconductivity~\cite{Norman2019similarities,RN78,RN42,RN63,RN41,wu2019robust,nomura2019formation,gao_twoband,RN66,RN68,L2020,zhang2019selfdoped,RN72,Jiang2019,RN76,Zhou2019,Werner2020,Hu2019}.

However, there is an important difference between infinite-layer
nickelate $R$NiO$_2$ ($R$ = Nd, La) and infinite-layer cuprate
CaCuO$_2$ in their electronic structures:
in infinite-layer cuprates, only a single Cu-$d_{x^2-y^2}$ band
crosses the Fermi level, while in infinite-layer nickelates, in
addition to Ni-$d_{x^2-y^2}$ band, another conduction band also
crosses the Fermi
level~\cite{lee2004infinite,wu2019robust,nomura2019formation,gao_twoband}.
First-principles calculations show that the other non-Ni conduction
band originates from rare-earth spacer
layers~\cite{lee2004infinite,wu2019robust,nomura2019formation,gao_twoband}.
Hepting \textit{et al}.~\cite{RN41} propose that itinerant electrons on rare-earth-$d$
orbitals may hybridize with Ni-$d_{x^2-y^2}$ orbital, rendering
$R$NiO$_2$ an ``oxide-intermetallic'' compound. But previous studies
find that the hybridization between Ni-$d_{x^2-y^2}$ and
rare-earth-$d$ orbitals is very weak~\cite{Jiang2019,  nomura2019formation, wu2019robust,gao_twoband}.  Therefore other
than the self-doping effect~\cite{zhang2019selfdoped}, infinite-layer
nickelates can still be considered as a promising analog of
infinite-layer cuprates~\cite{wu2019robust, Norman2019similarities}.
 
In this work, we combine density functional theory
(DFT)~\cite{Hohenberg-PR-1964, Kohn-PR-1965} and dynamical mean field
theory (DMFT)~\cite{Georges-RMP-1996, Kotliar-RMP-2006} to show that
the hybridization between Ni-$d_{x^2-y^2}$ orbital and itinerant
electrons in rare-earth spacer layers is substantially stronger than
previously thought. However, the largest source of hybridization comes
from an interstitial-$s$ orbital due to a large inter-cell
hopping. The hybridization with rare-earth-$d$ orbitals is weak, about
one order of magnitude smaller. We also find that weak-to-moderate
correlation effects on Ni lead to a charge transfer from
Ni-$d_{x^2-y^2}$ orbital to hybridization states, which provides more
itinerant electrons to couple to Ni-$d_{x^2-y^2}$ orbital.  In the
experimentally observed paramagnetic metallic state of $R$NiO$_2$, we
explicitly demonstrate that the coupling between
Ni-$d_{x^2-y^2}$ orbital and itinerant electrons screens the Ni local
moment, as in Kondo systems~\cite{Kondo1964, Wilson1983,
  Sawatzky2019}. Finally we find that the hybridization increases the
critical $U_{\textrm{Ni}}$ that is needed to induce long-range
magnetic ordering.

Our work provides the microscopic origin of
a substantial hybridization between Ni-$d_{x^2-y^2}$
orbital and itinerant electrons in $R$NiO$_2$, which leads to an
electronic structure that is distinct from that of CaCuO$_2$. As a
consequence of the hybridization, spins on
Ni-$d_{x^2-y^2}$ orbital are affected by itinerant electrons and the
physical property of $R$NiO$_2$ is changed. This implies that the
observed superconductivity in infinite-layer nickelates does not
emerge from a doped Mott insulator as in cuprates.

The computational details of our DFT and DMFT calculations can be
found in the Method section. For clarity, we study NdNiO$_2$ as a
representative of infinite-layer nickelates. The results of LaNiO$_2$
are very similar (see Supplementary Note 1 and Note 2 in the Supplementary
Information).

\section*{Results}

\subsection*{Electronic structure and interstitial-$s$ orbital}

\begin{figure}[!htb]
  \includegraphics[width=\textwidth]{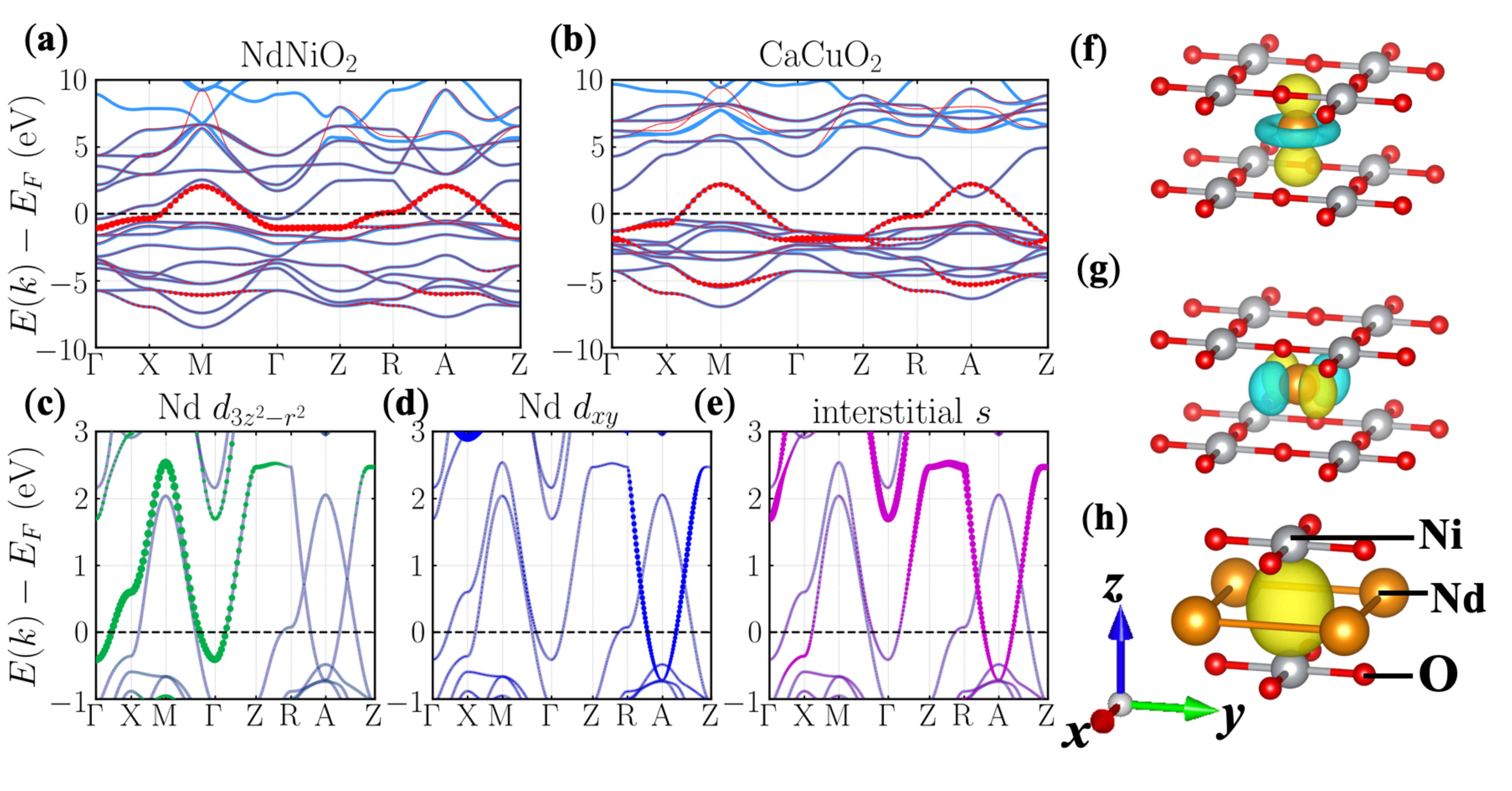}
  \caption{\textbf{Non-spin-polarized band structures calculated by density functional theory (DFT) and Wannier
      fitting}:  (\textbf{a-b}) DFT-calculated band structures and 17
    Wannier functions fitting of NdNiO$_2$ (\textbf{a}) and CaCuO$_2$
    (\textbf{b}). The thick blue lines are DFT-calculated bands and
    the red thin lines are bands reproduced by the Wannier functions.
    The red dots show the Wannier projection onto Ni-$d_{x^2-y^2}$ and
    Cu-$d_{x^2-y^2}$ orbitals, respectively. (\textbf{c-e}) Band
    structures reproduced by Wannier functions in an energy window
    close to the Fermi level.  The dots show the weights of Wannier
    projections onto Nd-$d_{3z^2-r^2}$ orbital (\textbf{c}),
    Nd-$d_{xy}$ orbital (\textbf{d}) and interstitial-$s$ orbital
    (\textbf{e}).  The coordinates of the high-symmetry points on the
    \textbf{k}-path are $\Gamma$(0,0,0)-$X$(0.5,0,0)-$M$(0.5,0.5,0)-$\Gamma$(0,0,0)-$Z$(0,0,0.5)-$R$(0.5,0,0.5)-$A$(0.5,0.5,0.5)-$Z$(0,0,0.5). The Fermi level $E_F$ (black dashed line) is shifted to zero
    energy. (\textbf{f-h}) An iso-value surface of the Wannier functions
    of Nd-$d_{3z^2-r^2}$ orbital (\textbf{f}),  Nd-$d_{xy}$ orbital (\textbf{g})
    and interstitial-$s$ orbital (\textbf{h}). The large
    orange atom is Nd, the gray atom is Ni and the small red atom is
    O.}
\label{fig:wannier}
\end{figure}

In Fig.~\ref{fig:wannier}(\textbf{a-b}), we show the
DFT-calculated band structure and Wannier function fitting of
NdNiO$_2$ and CaCuO$_2$ in the non-spin-polarized state,
respectively. We use altogether 17 Wannier projectors to fit the DFT
band structure: 5 Ni/Cu-$d$ orbitals, 5 Nd/Ca-$d$ orbitals, 3 O-$p$
orbitals (for each O atom) and an interstitial-$s$ orbital. The
interstitial-$s$ orbital is located at the position of the missing
apical oxygen. The importance of interstitial-$s$ orbitals has been
noticed in the study of electrides and infinite-layer
nickelates~\cite{nomura2019formation,hirayama2018electrides,matsuishi2003high}.
Our Wannier fitting exactly reproduces not only the band structure of
the entire transition-metal and oxygen $pd$ manifold, but also the
band structure of unoccupied states about 5 eV above the Fermi
level. In particular, the Ni/Cu-$d_{x^2-y^2}$ Wannier projector is
highlighted by red dots in Fig.~\ref{fig:wannier}(\textbf{a-b}). The
details of the Wannier fitting can be found in Supplementary Note 3 in
the Supplementary Information.
For both compounds, Ni/Cu-$d_{x^2-y^2}$ band crosses the
Fermi level. However, as we mentioned in the Introduction, in addition
to Ni-$d_{x^2-y^2}$ band, another conduction band also crosses the
Fermi level in NdNiO$_2$. Using Wannier analysis, we find that the
non-Ni conduction electron band is mainly composed of three orbitals:
Nd-$d_{3z^2-r^2}$, Nd-$d_{xy}$ and interstitial-$s$ orbitals. The
corresponding Wannier projectors are highlighted by dots in the panels of
Fig.~\ref{fig:wannier}(\textbf{c-e}). An iso-value surface
of the three Wannier functions (Nd-$d_{3z^2-r^2}$, Nd-$d_{xy}$ and
interstitial-$s$ orbitals) is explicitly shown in
Fig.~\ref{fig:wannier}(\textbf{f-h}). We note that interstitial-$s$
orbital is more delocalized than Nd-$d_{3z^2-r^2}$ and Nd-$d_{xy}$
orbitals. Because all these three orbitals
are located in the Nd spacer layer between adjacent NiO$_2$ planes, if
these three orbitals can hybridize with Ni-$d_{x^2-y^2}$ orbital, then
they will create a three-dimensional electronic structure, distinct
from that of CaCuO$_2$~\cite{RN41}.

\subsection*{Analysis of hybridization}

However, from symmetry consideration, \textit{within the same cell}
the hopping between Ni-$d_{x^2-y^2}$ orbital and any of those three
orbitals (Nd-$d_{3z^2-r^2}$, Nd-$d_{xy}$ and interstitial-$s$) is
exactly equal to zero~\cite{nomura2019formation}, which leads to the
conclusion that the hybridization between Ni-$d_{x^2-y^2}$ and
rare-earth-$d$ orbitals is weak~\cite{RN41,Jiang2019,  nomura2019formation}. While this conclusion is correct by itself,
the hybridization between Ni-$d_{x^2-y^2}$ and interstitial-$s$
orbital has been omitted in previous
studies~\cite{RN41,wu2019robust,gao_twoband,zhang2019selfdoped,
  Jiang2019, nomura2019formation}.  We find that due to a
large \textit{inter-cell} hopping, Ni-$d_{x^2-y^2}$ orbital hybridizes
with interstitial-$s$ orbital much more substantially than with
rare-earth-$d$ orbitals by about one order of magnitude.

\begin{figure}[!htb]
  \includegraphics[width=\textwidth]{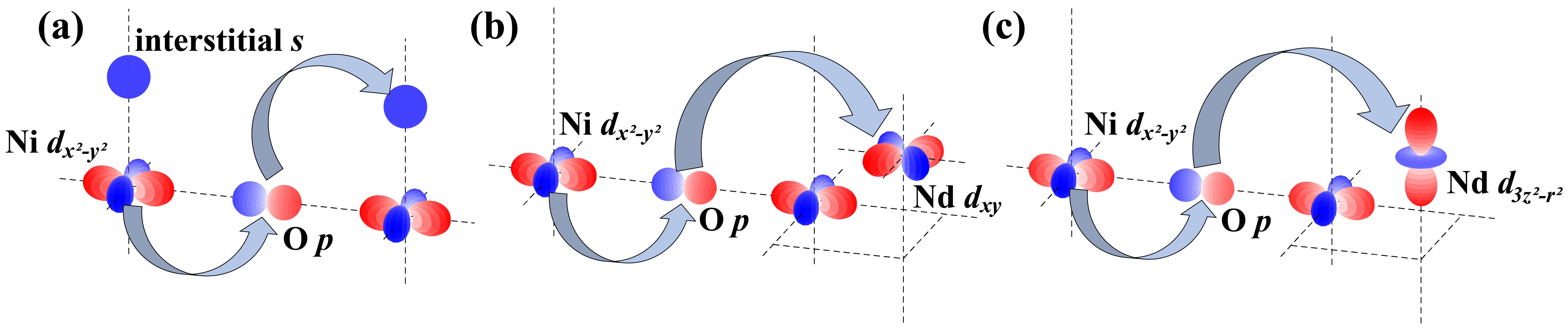}
  \caption{\textbf{Hybridization in NdNiO$_2$}: inter-cell hopping
    from Ni-$d_{x^2-y^2}$ to interstitial-$s$ orbital (\textbf{a}), to
    Nd-$d_{xy}$ orbital (\textbf{b}) and to Nd-$d_{3z^2-r^2}$ orbital
    (\textbf{c}) via one O-$p$ orbital.}
\label{fig:hopping}
\end{figure}

The direct inter-cell hopping between Ni-$d_{x^2-y^2}$ and any of the
three orbitals (Nd-$d_{3z^2-r^2}$, Nd-$d_{xy}$ and interstitial-$s$)
is negligibly small. The effective hopping is via O-$p$
orbitals. Fig.~\ref{fig:hopping} shows the inter-cell hopping between
Ni-$d_{x^2-y^2}$ orbital and the other three orbitals via one O-$p$
orbital. Among Nd-$d_{3z^2-r^2}$, Nd-$d_{xy}$ and interstitial-$s$
orbitals, we find that the largest effective hopping (via one O-$p$
orbital) is the one with interstitial-$s$ orbital (see
Table~\ref{tab:hops}).  The effective hopping between Ni-$d_{x^2-y^2}$
orbital and Nd-$d_{xy}$/$d_{3z^2-r^2}$ orbital is one order of
magnitude smaller because Nd atom is located at the corner of the
cell, which is further from the O atom than the interstitial site is.
Furthermore, the energy difference between interstitial-$s$ orbital
and O-$p$ orbital is about 1 eV smaller than that between
Nd-$d_{xy}$/$d_{3z^2-r^2}$ orbital and O-$p$ orbital (see
Table~\ref{tab:hops}). These two factors combined lead to the fact
that Ni-$d_{x^2-y^2}$ has a significant coupling with
interstitial-$s$ orbital, substantially stronger than that with Nd-$d$
orbitals. This challenges the previous picture that the hybridization
between Ni-$d_{x^2-y^2}$ orbital and itinerant electrons in the Nd spacer layer is weak~\cite{RN41,wu2019robust,gao_twoband,zhang2019selfdoped, Jiang2019, nomura2019formation}.

To further confirm that the hybridization is substantial, we
downfold the full band structure to a non-interacting model that is
based on the above four orbitals (Ni-$d_{x^2-y^2}$, Nd-$d_{3z^2-r^2}$,
Nd-$d_{xy}$ and interstitial-$s$ orbitals). Eq.~(\ref{eqn:100}) shows
the Wannier-based Hamiltonian $\langle \textbf{0} |H_0| \textbf{a}_1
\rangle = H_0(\textbf{a}_1)$ in the matrix form (not the usual
Hamiltonian $\langle \textbf{0} |H_0| \textbf{0} \rangle =
H_0(\textbf{0})$, $H_0(\textbf{0})$ is shown in Supplementary Note 3
in the Supplementary Information).
The important information is in the first row. The
largest hopping is the one between neighboring Ni-$d_{x^2-y^2}$ orbitals (this
is due to the $\sigma$ bond between Ni-$d_{x^2-y^2}$ and O-$p_x$/$p_y$
orbitals). However, the hopping between Ni-$d_{x^2-y2}$ and
interstitial-$s$ orbital is even comparable to the largest hopping. By
contrast, the hopping between Ni-$d_{x^2-y^2}$ orbital and Nd-$d_{xy}$
/$d_{3z^2-r^2}$ orbital is about one order of magnitude smaller than
the hopping between Ni-$d_{x^2-y^2}$ and interstitial-$s$ orbital,
which is consistent with the preceding analysis.
\begin{equation}
  \label{eqn:100}
    H_0(\mathbf{a}_1)=\bordermatrix{
            &d_{x^2-y^2} &  s      & d_{xy}  &  d_{3z^2-r^2}\cr
d_{x^2-y^2} & -0.37      & -0.22   &  0.03   & -0.02   \cr
s           & -0.22      &  -0.24  & 0.68    & 0.45    \cr
d_{xy}      &  0.03      &  0.68   & -0.08   & 0      \cr
d_{3z^2-r^2}     & -0.02      &  0.45   & 0       & -0.19}
\end{equation}

\subsection*{Charge transfer and screening of Ni local moment}

Since infinite-layer nickelates are correlated materials, next we
study correlation effects arising from Ni-$d_{x^2-y^2}$ orbital.  We
focus on whether the hybridization between Ni-$d_{x^2-y^2}$ orbital and
itinerant electrons in the rare-earth spacer layer may affect the
correlated properties of NdNiO$_2$, such as magnetism.

We use the above four orbitals (see Eq.~(\ref{eqn:100})) to build an
interacting model:
\begin{equation}
  \label{eqn:model}
\hat{H} = \sum_{\textbf{k},mm',\sigma}H_0(\textbf{k})_{mm'}\hat{c}^{\dagger}_{\textbf{k}m\sigma}\hat{c}_{\textbf{k}m'\sigma} + U_{\textrm{Ni}}\sum_{i}\hat{n}_{i\uparrow}\hat{n}_{i\downarrow} - \hat{V}_{dc}
\end{equation}
where $mm'$ labels different orbitals, $i$ labels Ni sites and
$\sigma$ labels spins. $\hat{n}_{i\sigma}$ is the occupancy operator
of Ni-$d_{x^2-y^2}$ orbital at site $i$ with spin $\sigma$ and the
onsite Coulomb repulsion is only applied on the Ni-$d_{x^2-y^2}$
orbital.  $H_0(\textbf{k})$ is the Fourier transform of the
Wannier-based Hamiltonian
$H_0(\textbf{R})$~\cite{Hansmann_heterostructure} and $\hat{V}_{dc}$
is the double counting potential. That we do not explicitly
include O-$p$ states in the model is justified by noting that in
NdNiO$_2$ O-$p$ states have much lower energy than Ni-$d$ states,
which is different from perovskite rare-earth nickelates and
charge-transfer-type cuprates~\cite{RN41,RN63}. In the model
Eq.~(\ref{eqn:model}), the Ni-$d_{x^2-y^2}$ orbital is the correlated
state while the other three orbitals (interstitial-$s$ and
Nd-$d_{3z^2-r^2}$/$d_{xy}$) are non-interacting, referred to as
hybridization states.

\begin{figure}[!htb]
  \includegraphics[width=\textwidth]{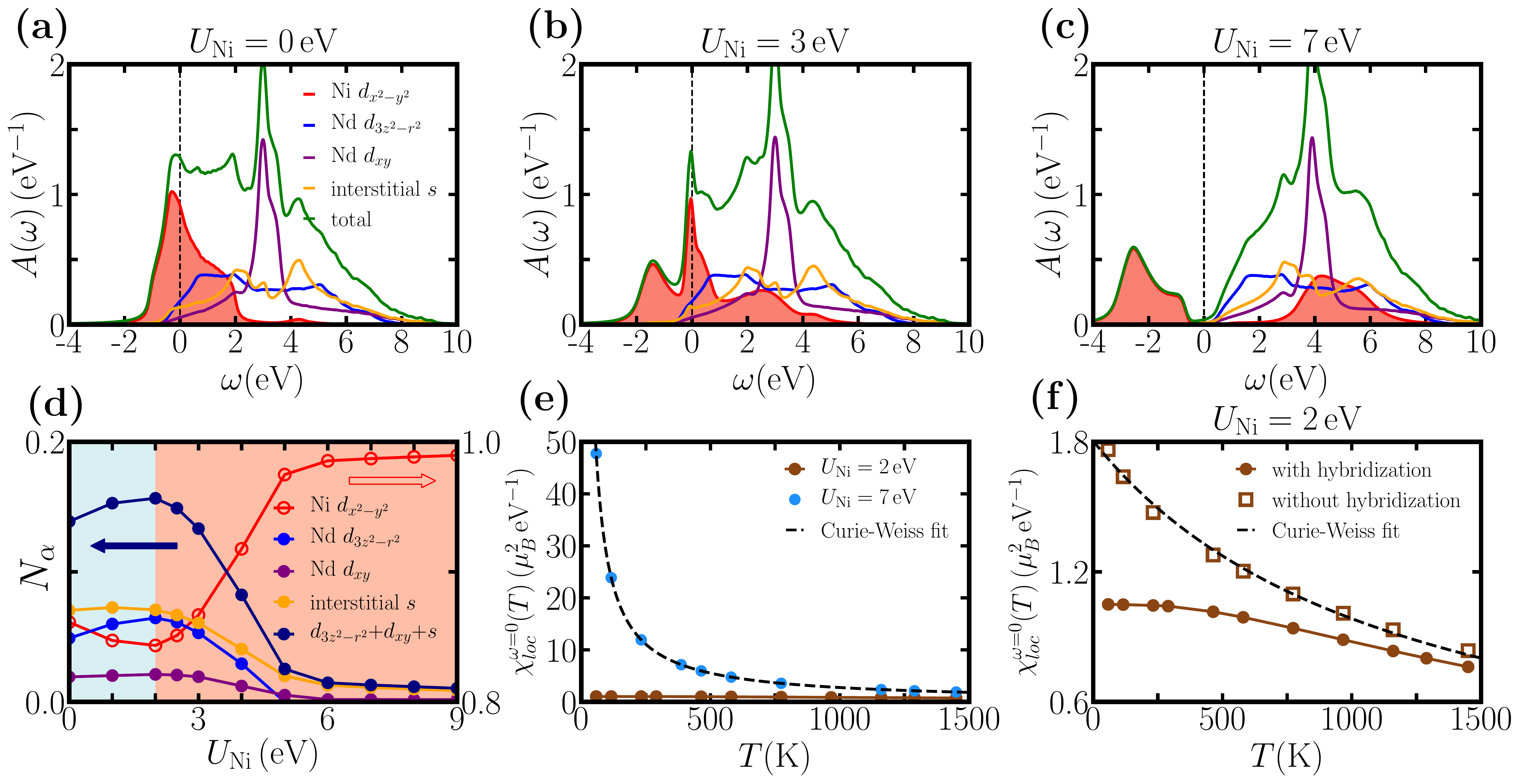}
  \caption{\textbf{Spectral function, orbital occupancy and local spin
      susceptibility of NdNiO$_2$ in the paramagnetic state}:
    (\textbf{a-c}) Spectral function of the 4-orbital interacting
    model Eq.~(\ref{eqn:model}) calculated by density functional theory plus dynamical mean field theory (DFT+DMFT) method     in the paramagnetic state at $T= 116$ K with $U_{\textrm{Ni}}$ = 0 eV
    (\textbf{a}), $U_{\textrm{Ni}}$ = 3 eV  (\textbf{b}) and 
    $U_{\textrm{Ni}}$ = 7 eV (\textbf{c}). $U_{\textrm{Ni}}$ is the Hubbard $U$ on each Ni atom. $A(\omega)$ is the frequency-dependent spectral function and $\omega$ represents the frequency. The Fermi level (black dashed line) is
    shifted to zero energy.  The red, blue, magenta, yellow and green
    curves are Ni-$d_{x^2-y^2}$ projected spectral function,
    Nd-$d_{3z^2-r^2}$ projected spectral function, Nd-$d_{xy}$
    projected spectral function, interstitial-$s$ projected spectral
    function and total spectral function, respectively. The
    Ni-$d_{x^2-y^2}$ projected spectral function is highlighted by red
    shades.  (\textbf{d}) Wannier function occupancy $N_{\alpha}$ as a
    function of $U_{\textrm{Ni}}$. Left axis is the occupancy of
    Nd-$d_{3z^2-r^2}$ (blue solid symbols), Nd-$d_{xy}$ (magenta solid
    symbols), interstitial-$s$ (yellow solid symbols) and their sum
    (dark blue solid symbols). Right axis is the occupancy of
    Ni-$d_{x^2-y^2}$ (red open symbols). The light blue and light red
    shades show the range of $U_{\textrm{Ni}}$ in which the total
    occupancy of hybridization states increases and decreases,
    respectively. (\textbf{e-f}) Local spin susceptibility
    $\chi^{\omega=0}_{\textrm{loc}}(T)$ of Ni-$d_{x^2-y^2}$ orbital as
    a function of tempreture $T$.
    (\textbf{e}) The blue symols are calculated using
    $U_{\textrm{Ni}}$ = 7 eV. The brown
    symbols are calculated using $U_{\textrm{Ni}}$ = 2 eV.  The black
    dashed line is a Curie-Weiss fitting.  (\textbf{f}) The solid
    symbols are the same as (\textbf{e}) in which the hybridization
    between itinerant electrons and Ni-$d_{x^2-y^2}$ orbital is
    ``turned on''. The open symbols are the local spin susceptibility $\chi^{\omega=0}_{\textrm{loc}}(T)$ recalculated at
    $U_{\textrm{Ni}}$ = 2 eV with the hybridization ``turned
    off''. The black dashed line is a Curie-Weiss fitting. }
\label{fig:phase}
\end{figure}

We perform dynamical mean field theory calculations on
Eq.~(\ref{eqn:model}).  We first study paramagnetic state
(paramagnetism is imposed in the calculations).
Fig.~\ref{fig:phase} (\textbf{a-c}) shows the spectral function with an
increasing $U_{\textrm{Ni}}$ on Ni-$d_{x^2-y^2}$ orbital. At
$U_{\textrm{Ni}}$ = 0 eV, the system
is metallic with all the four orbitals crossing the Fermi level (the
main contribution comes from Ni-$d_{x^2-y^2}$). As $U_{\textrm{Ni}}$
increases to 3 eV, a quasi-particle
peak is evident with the other three orbitals still crossing the Fermi
level. We find a critical $U_{\textrm{Ni}}$ of about 7 eV, where the
quasi-particle peak becomes completely suppressed and a Mott gap
emerges. As $U_{\textrm{Ni}}$
further increases to 9 eV (not shown in Fig.~\ref{fig:phase}), a clear
Mott gap of about 1 eV is opened. 

The presence of hybridization states means that there could be charge
transfer between correlated Ni-$d_{x^2-y^2}$ orbital and
interstitial-$s$/Nd-$d$ orbitals. We calculate the occupancy of each
Wannier function $N_{\alpha}$ and study correlation-driven charge
transfer in NdNiO$_2$. Fig.~\ref{fig:phase}(\textbf{d}) shows
$N_{\alpha}$ of each hybridization state and Ni-$d_{x^2-y^2}$ orbital
as well as the total occupancy of hybridization states as a function
of $U_{\textrm{Ni}}$.  We first note that at $U_{\textrm{Ni}}$ = 0,
the total occupany of hybridization states is 0.14, which is
significant. As $U_{\textrm{Ni}}$ becomes larger, the total occupany
of hybridization states first increases and then decreases.  This is
because when $U_{\textrm{Ni}}$ is small, the system is still metallic
with all the hybridization states crossing the Fermi level, while the
upper Hubbard band of Ni-$d_{x^2-y^2}$ orbital is just formed and
pushed to higher energy. This leads to charge transfer from
Ni-$d_{x^2-y^2}$ orbital to hybridization states, providing more
itinerant electrons to couple to Ni-$d_{x^2-y^2}$ orbital. However,
when $U_{\textrm{Ni}}$ is large, hybridization states are also pushed
above the Fermi level, which causes electron to transfer back to
Ni-$d_{x^2-y^2}$ orbital (in the lower Hubbard band). In the strong
$U_{\textrm{Ni}}$ limit where the Mott gap opens, itinerant electrons
in the Nd spacer layer disappear. Fig.~\ref{fig:phase}(\textbf{d}) also
shows that for all $U_{\textrm{Ni}}$ considered, the occupancy on
interstitial-$s$ orbital is always the largest among the three hybridization
states, confirming the importance of the
interstitial-$s$ orbital in infinite-layer nickelates. We
note that because we calculate the occupancy at finite temperatures,
even when the gap is opened, the
occupancy of hybridization states does not exactly become zero.

Because of the hybridization, we study possible
screening of Ni local magnetic moment by itinerant electrons.
We calculate local spin susceptibility of Ni-$d_{x^2-y^2}$ orbital:
\begin{equation}
  \label{eqn:chi}
  \chi^{\omega=0}_{\textrm{loc}}(T) = \int^{\beta}_0 \chi_{\textrm{loc}}(\tau)d\tau
  =\int^{\beta}_0g^2\langle S_z(\tau)S_z(0) \rangle  d\tau  
\end{equation}
where $S_z(\tau)$ is the local spin operator for Ni-$d_{x^2-y^2}$
orbital, at the imaginary time $\tau$. $g$ denotes the electron spin
gyromagnetic factor and $\beta= 1/(k_BT)$ is the inverse temperature.
Fig.~\ref{fig:phase}(\textbf{e}) shows
$\chi^{\omega=0}_{\textrm{loc}}(T)$ for two representative values of
$U_{\textrm{Ni}}$. The blue symbols are
$\chi^{\omega=0}_{\textrm{loc}}(T)$ for $U_{\textrm{Ni}} = 7$ eV when
the system becomes insulating. The local spin susceptibility nicely
fits to a Curie-Weiss behavior, as is shown by the black dashed line
in Fig.~\ref{fig:phase}(\textbf{e}). $\chi^{\omega=0}_{\textrm{loc}}(T)$
has a strong enhancement at low temperatures.
However, for $U_{\textrm{Ni}} = 2$ eV when the
system is metallic, we find a completely different
$\chi^{\omega=0}_{\textrm{loc}}(T)$. The local spin susceptibility has
very weak dependence on temperatures (see Fig.~\ref{fig:phase}(\textbf{f})
for the zoomin). In particular at low
temperatures ($T < 250$ K), $\chi^{\omega=0}_{\textrm{loc}}(T)$ reaches
a plateau. We note that the weak temperature dependence of
$\chi^{\omega=0}_{\textrm{loc}}(T)$ is consistent with the
experimental result of LaNiO$_2$ paramagnetic susceptibility
~\cite{hayward1999sodium}, in particular our simple
model calculations qualitatively reproduce the low-temperature plateau feature
that is observed in experiment~\cite{hayward1999sodium}. 

To explicitly understand how the hybridization between itinerant
electrons and Ni-$d_{x^2-y^2}$ orbital affects local spin
susceptibility, we perform a thought-experiment: we manually ``turn
off'' hybridization, i.e. for each $\textbf{R}$, we set $\langle
s|H_0(\textbf{R})| d_{x^2-y^2} \rangle= \langle d_{xy}
|H_0(\textbf{R})| d_{x^2-y^2} \rangle = \langle
d_{3z^2-r^2}|H_0(\textbf{R})| d_{x^2-y^2} \rangle = 0$. Then we
recalculate $\chi^{\omega=0}_{\textrm{loc}}(T)$ using the modified
Hamiltonian with $U_{\textrm{Ni}} = 2$ eV. The chemical
  potential is adjusted so that the total occupancy remains unchanged
  in the modified Hamiltonian. The two local spin susceptibilities
are compared in Fig.~\ref{fig:phase}(\textbf{f}). With hybridization,
$\chi^{\omega=0}_{\textrm{loc}}(T)$ saturates at low temperatures,
implying that $\mu_{\textrm{eff}}$ decreases or even vanishes with
lowering temperatures. However, without hybridization,
$\chi^{\omega=0}_{\textrm{loc}}(T)$ shows an evident enhancement at
low temperatures and a Curie-Weiss behavior is restored (black dashed
line).  This shows that in paramagnetic metallic NdNiO$_2$, the
hybridization between itinerant electrons and Ni-$d_{x^2-y^2}$ orbital
is substantial and as a consequence, it screens the
Ni local magnetic moment, as in Kondo systems~\cite{Kondo1964,
  Wilson1983,Sawatzky2019}.  Such a screening mechanism may be used to
explain the low-temperature upturn in the resistivity of NdNiO$_2$
observed in experiment~\cite{zhang2019selfdoped,
  li2019superconductivity}. We note that while we only
  fix the total occupancy by adjusting the chemical potential, the
  occupancy of Ni-$d_{x^2-y^2}$ orbital is almost the same in the
  original and modified models. In Fig.~\ref{fig:phase}(\textbf{f}),
  ``with hybridization'', Ni-$d_{x^2-y^2}$ occupancy is 0.84 and ``without
  hybridization'', Ni-$d_{x^2-y^2}$ occupancy is 0.83. This indicates that the
  screening of Ni moment is mainly due to the hybridization effects,
  while the change of Ni-$d_{x^2-y^2}$ occupancy (0.01$e$ per Ni)
  plays a secondary role.

\subsection*{Correlation strength and phase diagram}

\begin{figure}[!htb]
\includegraphics[width=\textwidth]{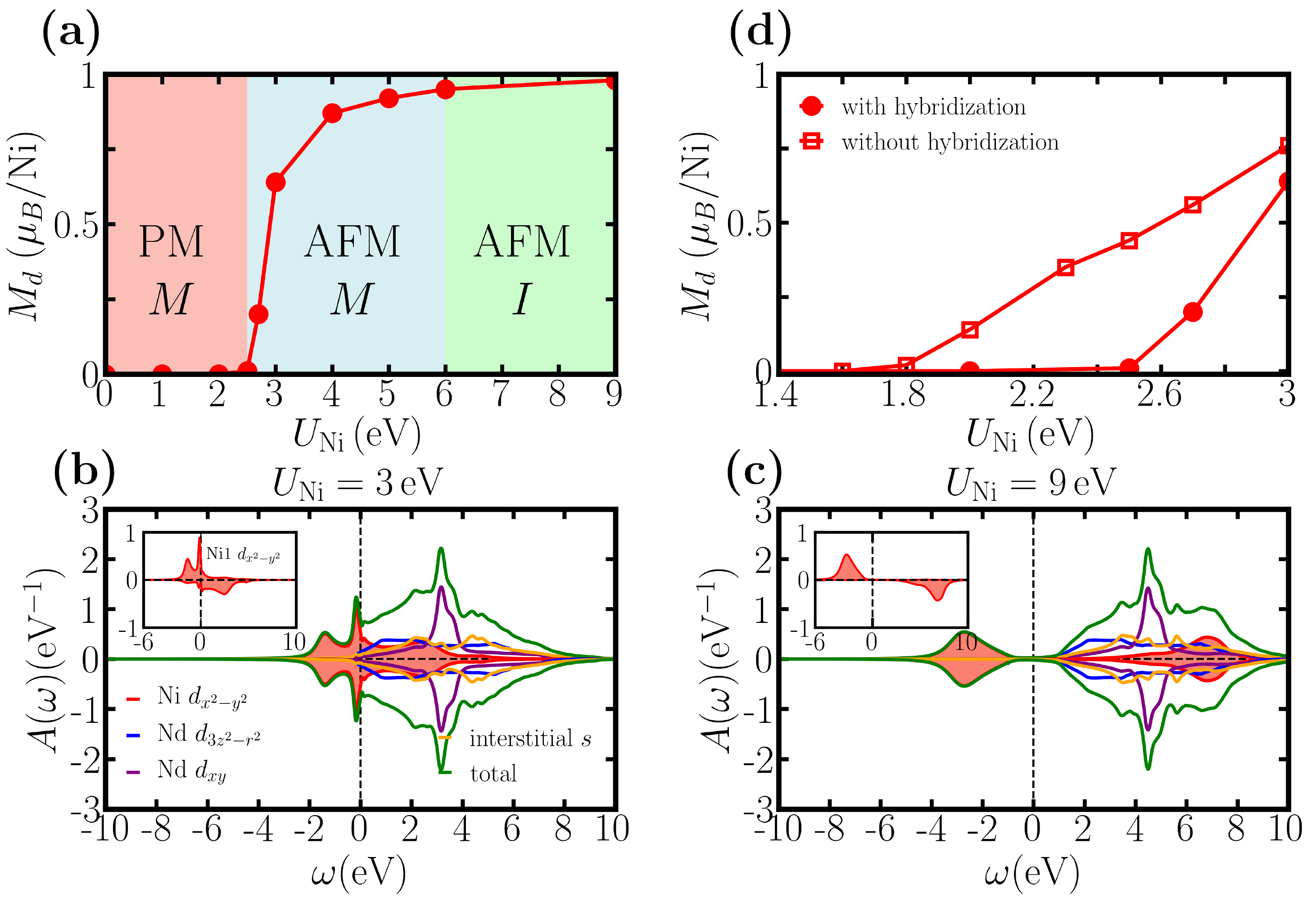}
\caption{\textbf{Phase diagram and antiferromagnetic spectral function
    of NdNiO$_2$}: (\textbf{a}) Phase diagram of NdNiO$_2$, calculated
  by using density functional theory plus dynamical mean field theory
  (DFT+DMFT) method based on the 4-orbital interacting model
  Eq.~(\ref{eqn:model}).  $M_d$ is the local moment on each Ni atom and
  $U_{\textrm{Ni}}$ is the Hubbard $U$ on each Ni atom.  `PM' means
  paramagnetic state; `AFM' means checkerboard antiferromagnetic
  state; italic `$M$' means `metallic'; italic `$I$' means
  `insulating'. (\textbf{b}) Spectral function of the 4-orbital
  interacting model Eq.~(\ref{eqn:model}) in the antiferromagnetic
  state with $U_{\textrm{Ni}} = 3$ eV. $A(\omega)$ is the
  frequency-dependent spectral function and $\omega$ represents the
  frequency. The states above (below) zero correspond to spin up
  (down). The Fermi level (vertical dashed line) is set at zero
  energy.  The red, blue, magenta, yellow and green curves represent
  Ni-$d_{x^2-y^2}$ projected spectral function, Nd-$d_{3z^2-r^2}$
  projected spectral function, Nd-$d_{xy}$ projected spectral
  function, interstitial-$s$ projected spectral function and total
  spectral function, respectively. The inset shows the spectral
  function of a \textit{single} Ni atom projected onto its
  $d_{x^2-y^2}$ orbital. (\textbf{c}) Same as (\textbf{b}) with
  $U_{\textrm{Ni}} = 9$ eV. (\textbf{d}) The solid symbols are the
  same as (\textbf{a}). The open symbols are local moment on each Ni
  atom recalculated with the hybridization ``turned off''.}
\label{fig:md}
\end{figure}

We estimate the correlation strength for NdNiO$_2$ by calculating its
phase diagram. We allow spin polarization in the DMFT
calculations and study both ferromagnetic and checkerboard
antiferromagnetic states. We find that ferromagnetic ordering can not
be stabilized up to $U_{\textrm{Ni}}$ = 9 eV. Checkerboard
antiferromagnetic state can emerge when $U_{\textrm{Ni}}$ exceeds 2.5
eV. The phase diagram is shown in Fig.~\ref{fig:md}(\textbf{a})
in which $M_d$ is the local magnetic moment on each Ni atom. $M_d$ is
zero until $U_{\textrm{Ni}} \simeq 2.5$ eV and then increases with
$U_{\textrm{Ni}}$ and finally saturates to 1 $\mu_B$/Ni which
corresponds to a $S=\frac{1}{2}$ state. We note that the critical
value of $U_{\textrm{Ni}}$ is model-dependent. If we include O-$p$
states and semi-core states, the critical value of $U_{\textrm{Ni}}$
will be substantially larger~\cite{Sakuma2013}.
The robust result here is that with
$U_{\textrm{Ni}}$ increasing, antiferromagnetic ordering occurs before
the metal-insulator transition. In the antiferromagnetic state, the
critical $U_{\textrm{Ni}}$ for the metal-insulator transition is about
6 eV, slightly smaller than that in the paramagnetic phase. The
spectral function of antiferromagnetic metallic and insulating states
is shown in Fig.~\ref{fig:md} (\textbf{b}) and (\textbf{c}), respectively.
Experimentally long-range magnetic orderings are not
observed in NdNiO$_2$~\cite{hayward2003NdNiO2}.
The calculated phase diagram means that NdNiO$_2$ can only be in a
paramagnetic metallic state (instead of a paramagnetic insulating
state), in which the hybridization between Ni-$d_{x^2-y^2}$ and itinerant
electrons screens the Ni local magnetic moment.
We note that using our model Eq.~(\ref{eqn:model}),
the calculated phase boundary
indicates that Ni correlation strength is moderate in NdNiO$_2$ with
$U_{\textrm{Ni}}/t_{dd}$ less than 7 ($t_{dd}$ is the effective
hopping between nearest-neighbor Ni-$d_{x^2-y^2}$ due to $\sigma_{pd}$
bond). This contrasts with the parent compounds of superconducting
cuprates which are antiferromagnetic insulators and are described by
an effective single-orbital Hubbard model with a
larger correlation strength ($U_{\textrm{Ni}}/t_{dd} =  8\sim20$)
~\cite{tokura1990cu, NatPhys2008, PhysRevB.80.054501,
  NatPhys2010}. Finally, we perform a self-consistent check
on the hybridization. When the system is metallic, the hybridization
between itinerant electrons and Ni-$d_{x^2-y^2}$ orbital screens
the spin on Ni site and reduces the local spin susceptibility
$\chi^{\omega=0}_{\textrm{loc}}(T)$ in the paramagnetic phase. This
implies that once we allow antiferromagnetic ordering, a smaller
critical $U_{\textrm{Ni}}$ may be needed to induce magnetism. 
To test that, we recalculate the phase diagram using the modified
Hamiltonian with the hybridization manually ``turned off''. The chemical potential is adjusted in the modified model so that the total
occupancy remains unchanged.
Fig.~\ref{fig:md}(\textbf{d}) shows that without the hybridization,
the Ni magnetic moment increases and the antiferromagnetic phase is
expanded with the critical $U_{\textrm{Ni}}$ reduced to 1.8 eV
($U_{\textrm{Ni}}/t_{dd} \simeq 5$). This shows
that the coupling to the conducting electrons affects Ni spins and
changes the magnetic property of NdNiO$_2$~\cite{Sawatzky2019}.

\section*{Discussion}

\begin{figure}[!htb]
\includegraphics[width=0.8\textwidth]{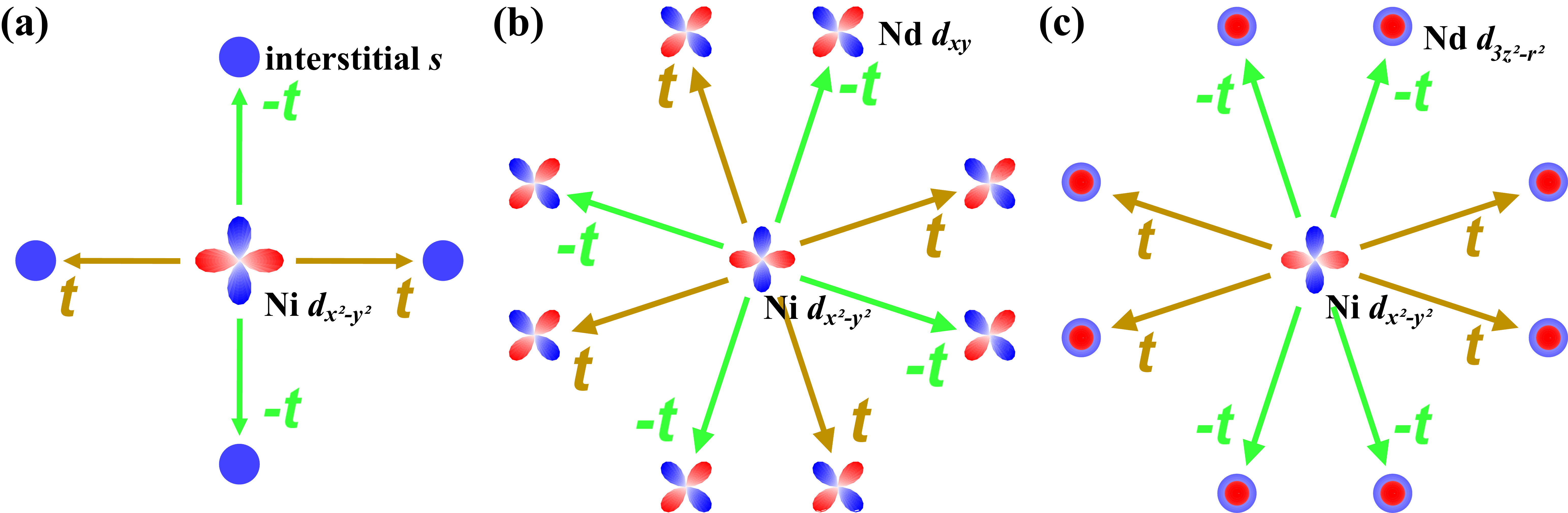}
\caption{\label{fig:symmetry} \textbf{Symmetry of hybridization in NdNiO$_2$}:
  intercell hopping from
  Ni-$d_{x^2-y^2}$ orbital to intersitial-$s$
  orbital (\textbf{a}), to Nd-$d_{xy}$ orbital (\textbf{b}) and to
  Nd-$d_{3z^2-r^2}$ orbital (\textbf{c}). All the hoppings here are between
  second nearest neighbors. Brown and green arrows represent
  positive and negative hoppings, respectively. Note this is a
  top view. Nd spacer layer and NiO$_2$ layer are not in the same
  plane.}
\end{figure}

Our minimal model Eq.~(\ref{eqn:model}) is different
from the standard Hubbard model (single-orbital, two-dimensional square
lattice and half filling) due to the presence of
hybridization. It is also different from a
standard periodic Anderson model in that 1) the correlated orbital is a
$3d$-orbital with a strong dispersion instead of a $4f$ or $5f$
orbital whose dispersion is usually neglected ~\cite{RN41,
  bulla2008numerical, PhysRevLett.85.373}; 2) the
hybridization of Ni-$d_{x^2-y^2}$ with the three non-interacting
orbitals is all inter-cell rather than onsite and anisotropic with
different types of symmetries, which may influence the symmetry of the
superconducting order parameter in the ground
state~\cite{sarasua2003superconductivity}. Fig.~\ref{fig:symmetry}
explicitly shows the symmetry of hybridization. The dominant hybridization
of Ni-$d_{x^2-y^2}$ orbital, the one with interstitial-$s$ orbital,
has $d_{x^2-y^2}$ symmetry.
Secondarily, the hybridization of
Ni-$d_{x^2-y^2}$ with Nd-$d_{xy}$ and Nd-$d_{3z^2-r^2}$ orbitals has
$g_{xy(x^2-y^2)}$ and $d_{x^2-y^2}$ symmetries, respectively~\cite{dresselhauss2007group}. 

$d$-wave superconducting states can be stabilized in the doped
single-orbital Hubbard model from sophisticated many-body
calculations~\cite{zheng2017stripe, Gull2013, Maier2005,
  Halboth2000}. However, the hybridization between
correlated Ni-$d_{x^2-y^2}$ orbital and itinerant electrons
fundamentally changes the electronic structure of a single-orbital
Hubbard model, in particular when the system is metallic. This
probably creates a condition unfavorable for superconductivity
~\cite{sarasua2003superconductivity}, implying that new mechanisms
such as interface charge transfer, strain engineering, etc. are needed
to fully explain the phenomena observed in infinite-layer nickelates
~\cite{li2019superconductivity}.

Before we conclude, we briefly discuss other modellings for $R$NiO$_2$ ($R$ = La, Nd). In literature, some
  models focus on low-energy physics and include only states that are close to the Fermi level; others include more states which reproduce the electronic band structure within a large energy window around the Fermi level. Kitatani \textit{et al.}~\cite{Held2020} propose that $R$NiO$_2$ can be described by the one-band Hubbard model (Ni-$d_{x^2-y^2}$ orbital) with an additional electron reservoir, which is used to directly estimate the superconducting transition temperature. Hepting \textit{et al.}~\cite{RN41} construct a two-orbital
  model using Ni-$d_{x^2-y^2}$ orbital and a $R$-$d_{3z^2-r^2}$-like orbital. Such a model is used to study hybridization effects between Ni-$d_{x^2-y^2}$ orbital and rare-earth $R$-$d$ orbitals. Zhang \textit{et al.}~\cite{RN72}, Werner \textit{et al.}~\cite{Werner2020} and Hu \textit{et al.}~\cite{Hu2019} study a different type of two-orbital models which consist of two Ni-$d$ orbitals. Hu \textit{et al.}~\cite{Hu2019} include Ni-$d_{x^2-y^2}$ and Ni-$d_{xy}$ orbitals, while Zhang \textit{et al.} and Werner \textit{et al.}~\cite{Werner2020, RN72} include  Ni-$d_{x^2-y^2}$ and Ni-$d_{3z^2-r^2}$ orbitals. This type of two-orbital model aims to study the possibility of high-spin $S=1$  doublon when the system is hole doped. Wu \textit{et al.}~\cite{wu2019robust} and Nomura \textit{et al.}~\cite{nomura2019formation} study three-orbital   models. Wu \textit{et al.}~\cite{wu2019robust} include Ni-$d_{xy}$ orbital,   $R$-$d_{xy}$ orbital and $R$-$d_{3z^2-r^2}$ orbital. This model is  further used to calculate the spin susceptibility and to estimate the superconducting transition temperature.  Nomura \textit{et al.}~\cite{nomura2019formation} compare two choices of orbitals:
  one is Ni-$d_{xy}$-orbital, $R$-$d_{3z^2-r^2}$ orbital and interstitial-$s$; and the other is Ni-$d_{xy}$-orbital,
  $R$-$d_{3z^2-r^2}$ orbital and $R$-$d_{xy}$. The model is used to study the screening effects on the Hubbard $U$ of Ni-$d_{x^2-y^2}$ orbital. Gao \textit{et al.}~\cite{gao_twoband} construct a general four-orbital model $B_{1g}@1a \bigoplus A_{1g}@1b$ which consists of two Ni-$d$ orbitals and two $R$-$d$ orbitals. The model is used to study the
  topological property of the Fermi surface. Jiang \textit{et al.}~\cite{Jiang2019}
  use a tight-binding model that consists of 5 Ni-$d$ orbitals and 5
  $R$-$d$ orbitals to comprehensively study the hybridization
  effects between Ni-$d$ and $R$-$d$ orbitals; Jiang \textit{et al.} also highlight the importance of
  Nd-$f$ orbitals in the electronic structure of NdNiO$_2$.
  Botana \textit{et al.}~\cite{Norman2019similarities}, Lechermann~\cite{L2020} and Karp \textit{et al.}~\cite{andypaper} consider
  more orbitals (including Nd-$d$, Ni-$d$ and O-$p$ states) in the modelling
  of NdNiO$_2$ with the interaction
  applied to Ni-$d$ orbitals and make a comparison to infinite-layer
  cuprates. Botana \textit{et al.}~\cite{Norman2019similarities} extract
  longer-range hopping parameters and the $e_g$ energy splitting.
  Lechermann~\cite{L2020} studies hybridization and doping
  effects. Karp \textit{et al.}~\cite{andypaper} calculate the phase diagram and
  estimates the magnetic transition temperature.
  
\section*{Conclusion}

In summary, we use first-principles calculations to study the
electronic structure of the parent superconducting material $R$NiO$_2$
($R$ = Nd, La). We find that the hybridization between
Ni-$d_{x^2-y^2}$ orbital and itinerant electrons is substantially
stronger than previously thought. The dominant hybridization comes
from an interstitial-$s$ orbital due to a large inter-cell
hopping, while the hybridization with rare-earth-$d$ orbitals is one order
of magnitude weaker. Weak-to-moderate
correlation effects on Ni cause electrons to
transfer from Ni-$d_{x^2-y^2}$ orbital to the hybridization states, which
provides more itinerant electrons in the rare-earth spacer layer to
couple to correlated Ni-$d$ orbital.
Further increasing correlation strength leads to a reverse
charge transfer, antiferromagnetism on Ni sites and eventually a metal-insulator
transition.
In the experimentally observed paramagnetic metallic state of $R$NiO$_2$,
we find that the strong coupling
between Ni-$d_{x^2-y^2}$ and itinerant electrons screens
the Ni local moment, as in Kondo systems.
We also find that the hybridization increases the critical $U_{\textrm{Ni}}$
that is needed to induce long-range magnetic ordering.
Our work shows that the electronic structure of $R$NiO$_2$ is
fundamentally different from that of CaCuO$_2$, which implies that
the observed superconductivity in infinite-layer nickelates does not emerge
from a doped Mott insulator as in cuprates.

\section*{Methods}

We perform first-principles calculations using density functional
theory (DFT)~\cite{Hohenberg-PR-1964, Kohn-PR-1965},
maximally localized Wannier functions (MLWF) to construct the non-interacting
tight-binding models~\cite{Marzari2012} and dynamical
mean field theory (DMFT)~\cite{Georges-RMP-1996, Kotliar-RMP-2006} to
solve the interacting models.

\subsection*{DFT calculations}

The DFT method is implemented in the Vienna ab initio simulation package
(VASP) code~\cite{Kresse1996} with the projector augmented wave (PAW)
method~\cite{kresse1999}. The Perdew-Burke-Ernzerhof
(PBE)~\cite{perdew1996} functional is used as the exchange-correlation
functional in DFT calculations. The Nd-$4f$ orbitals are treated as core
states in the pseudopotential. We use an energy cutoff of 600 eV and
sample the Brillouin zone by using $\Gamma$-centered \textbf{k}-mesh of
$16\times16\times16$. The crystal structure is fully relaxed with an energy
convergence criterion of $10^{-6}$ eV, force convergence
criterion of 0.01 eV/\AA~and strain convergence of 0.1 kBar.
The DFT-optimized crystal structures are in excellent agreement with the
experimental structures, as shown in our Supplementary Note 1.
To describe the checkerboard antiferromagnetic ordering, we expand the cell
to a $\sqrt{2}\times\sqrt{2}\times1$ supercell. The corresponding
Brillouin zone is sampled by using a $\Gamma$-centered \textbf{k}-mesh of
$12\times12\times16$. 

\subsection*{MLWF calculations}
We use maximally localized Wannier functions~\cite{Marzari2012}, as
implemented in Wannier90 code~\cite{mostofi2008wannier90} to fit the
DFT-calculated band structure and build an ab initio
tight-binding model which includes onsite energies and hopping
parameters for each Wannier function. We use two sets of Wannier
functions to do the fitting. One set uses 17 Wannier functions to
exactly reproduce the band structure of entire transition-metal and
oxygen $pd$ manifold as well as the unoccupied states that are a few
eV above the Fermi level. The other set uses 4 Wannier functions to
reproduce the band structure close to the Fermi level. The second
tight-binding Hamiltonian is used to study correlation effects when
onsite interactions are included on Ni-$d_{x^2-y^2}$ orbital.

\subsection*{DMFT calculations}
We use DMFT method to calculate the 4-orbital interacting model,
which includes a correlated Ni-$d_{x^2-y^2}$ orbital and three
non-interacting orbitals (interstitial-$s$, Nd-$d_{xy}$ and
Nd-$d_{3z^2-r^2}$). We also cross-check the results using a 17-orbital
interacting model which includes five Ni-$d$, five Nd-$d$, six O-$p$ and
one interstitial-$s$ orbital (the results of the 17-orbital model
are shown in Supplementary Note 4 of the Supplementary Information).
DMFT maps the interacting lattice Hamiltonian
onto an auxiliary impurity problem which is solved using the
continuous-time quantum Monte Carlo algorithm based on hybridization
expansion~\cite{Werner2006, Gull2011}. The impurity solver is
developed by K. Haule~\cite{Haule2007}. For each DMFT iteration, a
total of 1 billion Monte Carlo samples are collected to converge the
impurity Green function and self energy. We set the temperature to be
116 K. We check all the key results at a lower temperature of 58 K and
no significant difference is found.
The interaction strength $U_{\textrm{Ni}}$ is treated as a
parameter. We calculate both paramagnetic and magnetically ordered
states. For magnetically ordered states, we consider ferromagnetic
ordering and checkerboard antiferromagnetic ordering. For checkerboard
antiferromagnetic ordering calculation, we double the cell and the
non-interacting Hamiltonian is $8\times 8$. We introduce
formally two effective impurity models and use the symmetry that
electrons at one impurity site are equivalent to the electrons on
the other with opposite spins. The DMFT self-consistent condition
involves the self-energies of both spins.

To obtain the spectral functions, the imaginary axis self energy is
continued to the real axis using the maximum entropy
method~\cite{Silver1990}. Then the real axis local Green function is
calculated using the Dyson equation and the spectral function is
obtained following:

\begin{equation}
\label{eqn:spectral} A_m(\omega) = -\frac{1}{\pi}\textrm{Im} G^{\textrm{loc}}_m(\omega) = -\frac{1}{\pi}\textrm{Im}\left(\sum_{\textbf{k}}\frac{1}{(\omega+\mu)\mathbf{1}-H_0(\textbf{k})-\Sigma(\omega)+V_{dc}}\right)_{mm}
\end{equation}
where $m$ is the label of a Wannier function. $\mathbf{1}$ is an
identity matrix, $H_0(\textbf{k})$ is the Fourier transform of the
Wannier-based Hamiltonian
$H_0(\textbf{R})$. $\Sigma(\omega)$ is the self-energy,
understood as a diagonal matrix only with nonzero entries on the
correlated orbitals. $\mu$ is the chemical potential. $V_{dc}$ is the
fully localized limit (FLL) double counting potential, which is
defined as~\cite{Czyzyk1994}:

\begin{equation}
\label{eqn:dc} V_{dc} = U\left(N_d - \frac{1}{2}\right)
\end{equation}
where $N_d$ is the $d$ occupancy of a correlated site. Here the Hund's
$J$ term vanishes because we have a single correlated orbital
Ni-$d_{x^2-y^2}$ in the model. A $40\times40\times 40$ $k$-point mesh
is used to converge the spectral function.
We note that double counting correction affects
the energy separation between Ni-$d_{x^2-y^2}$ and
Nd-$d$/interstitial-$s$ orbitals. However because the charge transfer
is small (around 0.1$e$ per Ni), the effects from the double counting
correction are weak in the 4-orbital model, compared to those in the
$p$-$d$ model in which double counting correction becomes much more
important~\cite{Hyowon2014a}. That is because
O-$p$ states are included in the $p$-$d$ model. The double counting
correction affects the $p$-$d$ energy separation and thus the charge
transfer between metal-$d$ and oxygen-$p$ orbitals, which can be as
large as 1$e$ per metal atom for late transition metal oxides such as
rare-earth nickelates~\cite{Hyowon2014a}.

\section*{D\lowercase{ata availability}}
The data that support the findings of this study are available from
the corresponding author upon reasonable request.

\section*{C\lowercase{ode availability}}

The electronic structure calculations were performed using the proprietary
code VASP~\cite{Kresse1996}, the open-source code Wannier90
~\cite{mostofi2008wannier90} and the open-source impurity solver
implemented by Kristjan Haule at Rutgers University
(http://hauleweb.rutgers.edu/tutorials/).
Both Wannier90 and Haule's impurity solver are freely
distributed on academic use under the Massachusetts Institute of Technology
(MIT) License.

\clearpage
\newpage

\bibliographystyle{naturemag}
%\bibliography{Endnote}

\clearpage
\newpage

\section*{A\lowercase{CKNOWLEDGEMENTS}}
We thank Danfeng Li, Jean-Marc Triscone, Andrew
Millis, Dong Luan and Qiang Zhang for useful discussions. H. Chen is
supported by the National Natural Science Foundation of China under
project number 11774236 and NYU University Research Challenge Fund.
Y. Gu and J. Hu are supported by the Ministry of Science and
Technology of China 973 program (Grant No. 2015CB921300,
No.~2017YFA0303100, No. 2017YFA0302900), National Science Foundation
of China (Grant No. NSFC-11334012), the Strategic Priority Research
Program of CAS (Grant No. XDB07000000) and High-performance Computing
Platform of Peking University. Computational resources are provided by
High-performance Computing Platform of Peking University, NYU
High-performance computing at New York, Abu Dhabi and Shanghai
campuses.

\section*{A\lowercase{UTHOR CONTRIBUTIONS}}
H.C. conceived the project. Y.G. and H.C. performed the calculations.
S.Z. and X.W. analyzed data. J. H. participated in the discussion.
H.C. and Y.G. wrote the manuscript. All the authors commented on the
paper.

\vspace{0.5cm}

\textbf{Supplementary information} accompanies the paper on the
\textit{Communications Physics} website (DOI).

\textbf{Competing interests}: The authors declare no competing interests.

\clearpage
\newpage

\clearpage
\begin{table}[!ht]
  \caption{\label{tab:hops} \textbf{Hopping and energy difference between
      different orbitals of NdNiO$_2$}:
    the hopping $t$ and energy difference
  $\Delta$ between the five relevant orbitals of NdNiO$_2$ shown in
  Fig.~\ref{fig:hopping}: $d_{x^2-y^2}$ is the Ni-$d_{x^2-y^2}$
  orbital; $p$ is the O-$p$ orbital;  $d_{3z^2-r^2}$ is the
  Nd-$d_{3z^2-r^2}$ orbital;  $d_{xy}$ is the Nd-$d_{xy}$ orbital;
  $s$ is the interstitial-$s$ orbital.  The hopping and energy
  difference are obtained from 17 Wannier functions fitting. The unit
  is eV.}
\begin{ruledtabular}
\begin{tabular}{cccc}
  %\hline\hline
$t_{pd_{x^2-y^2}}$  & $t_{ps}$ & $t_{pd_{xy}}$  & $t_{pd_{3z^2-r^2}}$  \\
  1.31  &  -0.67  &  -0.06  & -0.03   \\
  \hline
  $\Delta_{d_{x^2-y^2}p}$  & $\Delta_{sp}$  & $\Delta_{d_{xy}p}$  & $\Delta_{d_{3z^2-r^2}p}$ \\
  3.98  &  6.74  & 7.64  & 7.53\\
  %\hline\hline
\end{tabular}
\end{ruledtabular}
\end{table}

\vspace{0.5cm}

\end{document}